\documentclass[twoside,twocolumn,9pt]{article}
\usepackage{extsizes}
\usepackage[super,sort&compress,comma]{natbib} 
\usepackage[version=3]{mhchem}
\usepackage[left=1.5cm, right=1.5cm, top=1.785cm, bottom=2.0cm]{geometry}
\usepackage{balance}
\usepackage{sectsty}
\usepackage{graphicx} 
\usepackage{lastpage}
\usepackage[format=plain,justification=justified,singlelinecheck=false,font={stretch=1.125,small,sf},labelfont=bf,labelsep=space]{caption}
\usepackage{float}
\usepackage{fancyhdr}
\usepackage{fnpos}
\usepackage[english]{babel}
\addto{\captionsenglish}{%
  
}
\usepackage{array}
\usepackage{droidsans}
\usepackage{charter}
\usepackage[T1]{fontenc}
\usepackage[usenames,dvipsnames]{xcolor}
\usepackage{setspace}
\usepackage[compact]{titlesec}
\usepackage{hyperref}

\begin{document}

\thispagestyle{plain}

\makeatother

\makeatletter 
\renewcommand\@biblabel[1]{#1}            
\renewcommand\@makefntext[1]%
{\noindent\makebox[0pt][r]{\@thefnmark\,}#1}


\twocolumn[
  \begin{@twocolumnfalse}

\vspace{1em}
\sffamily

{\textbf{\large{Symmetry-breaking, motion and bistability of active drops through polarization-surface coupling}}} \\

\large{Fenna Stegemerten,$^{\ast}$ Karin John,\textit{$^{a}$} and Uwe Thiele$^{\ast}$\textit{$^{b}$}$^{\ddag}$} \\

\normalsize{Cell crawling crucially depends on the collective dynamics of the acto-myosin cytoskeleton. However, it remains an open question to what extent cell polarization and persistent motion depend on continuous regulatory mechanisms \textbf{and} autonomous physical mechanisms. Experiments on cell fragments and theoretical considerations for active polar liquids have highlighted that physical mechanisms induce motility through splay and bend configurations in a nematic director field.  

Here, we employ a simple model, derived from basic thermodynamic principles, for active polar free-surface droplets to identify a different mechanism of motility. Namely, active stresses drive drop motion through spatial variations of polarization strength. This robustly induces parity-symmetry breaking and motility even for liquid ridges (2D drops) and adds to splay- and bend-driven pumping in 3D geometries.

Intriguingly, then, stable polar moving and axisymmetric resting states may coexist, reminiscent of the interconversion of moving and resting keratocytes by external stimuli. The identified additional motility mode originates from a competition between the elastic bulk energy and the polarity control exerted by the drop surface. As it already breaks parity-symmetry for passive drops, the resulting back-forth asymmetry enables active stresses to effectively pump liquid and drop motion ensues.} \\

\small{The published version of this preprint can be found under DOI:\url{10.1039/D2SM00648K}.}
 \end{@twocolumnfalse} \vspace{0.6cm}

  ]


\section*{}
\vspace{-1cm}


\footnotetext{$^{\ast}$~Institut f\"ur Theoretische Physik, Westf{\"a}lische Wilhelms-Universit\"at M\"unster, Wilhelm-Klemm-Str. 9, 48149 M\"unster, Germany}
\footnotetext{\textit{$^{a}$~Universit\'{e} Grenoble-Alpes, CNRS, Laboratoire Interdisciplinaire de Physique 38000 Grenoble, France }}
\footnotetext{\textit{$^{b}$~Center for Nonlinear Science (CeNoS), Westf{\"a}lische Wilhelms-Universit\"at M\"unster, Corrensstr.\ 2, 48149 M\"unster, Germany}}
\footnotetext{$^{\ddag}$~ORCID ID: 0000-0001-7989-9271, e-mail: u.thiele@uni-muenster.de, homepage: http://www.uwethiele.de}


\newcommand{\lowt}[1]{_{\text{#1}}}
\newcommand{\hight}[1]{^{\text{\tiny{#1}}}}


\section{Introduction}
Active media far from thermodynamic equilibrium display rich dynamic phenomena generated by active stress and self-propulsion \cite{GWSS2020jp,MJR+2013rmp}.   
From a reductionist's perspective, bacterial colonies, aggregates of epithelial cells or the cytoplasm of eukaryotic cells \cite{WTY+2018bj,DGN+2011potnaos, DDB2012sm,MDFYD2021natphys,VeSB1999cb} form a special class, where a thin free-surface layer of active matter is in contact with a solid substrate.
 There, active forces in the bulk related to nematic or polar order compete with interfacial forces acting at the free surface and the solid substrate. These autonomous physical mechanisms provide the backdrop for biological regulation pathways, and their interplay gives rise to experimental phenomena, such as active spreading and motility \cite{PAB+2019np,MGB+2017dc,SCD+2012n}.

Given, the complex nature of biological and biochemical systems, theoretical modeling of simplified or effective systems has proven to be a valuable tool to dissect the prevailing mechanism. Within the context of active drops as models for symmetry-breaking and motility in eukaryotic cells and cell fragments, many studies are limited to strictly planar geometries \cite{ZiSA2012jrsi,GD2014prl,ZiAr2016ncm,MWP2015jotrsi,TMC2012potnaos}, neglecting the direction perpendicular to the substrate and, in consequence, the physics of the curved free surface. In contrast, fully three-dimensional (3D) hydrodynamic models are rare and the full model behavior is costly to analyze \cite{TTMC2015nc}. 
Alternatively, long-wave approximations \cite{ODB1997rmp} are employed to derive thin-film models for shallow drops of active polar liquids  \cite{SR2009prl,JoRa2012jfm,KA2015pre, KMW2018potrsa,LoEL2019prl,AuET2020sm,WaQX2021sm,KA2015pre} and of passive nematic liquid crystals \cite{BeCu2001pf,LCA+2013pf,LKT+2013jofm}.
The latter approach may be extended to the active case by endowing the nematic order with an active stress and including evolution equations for the nematic order parameter \cite{KMW2018potrsa}.
One striking result of the above cited theoretical approaches is that active contractile stresses related to nematic order are sufficient to induce surface waves \cite{SR2009prl} and drop (cell) motion \cite{TMC2012potnaos,TTMC2015nc,KA2015pre,LoEL2019prl}. Indeed, fluid motion can be induced by spatial variations in the director field (splay, bend) either normal \cite{LoEL2019prl} or tangential to the substrate \cite{TMC2012potnaos,KA2015pre,SR2009prl}. In the latter case, drop motion requires a polar coupling between the liquid surface and the director field. 
The current consensus is that self-propulsion is not required for persistent motion as long as splay and bend configurations in the director field can be maintained. However, the onset of motion and the origin of the hysteresis observed experimentally in the keratocyte system \cite{VeSB1999cb} have not been studied in detail within this context.
Here, we provide a detailed analysis of the mechanism of motility in active polar droplets with a free surface as a simple physical model for cellular motility, reducing the biological complexity to effective parameters and variables.
Thereby we focus on the effect of coupling the bulk polarization to the drop's free surface shape, a crude realization of the idea, that the eukaryotic cell membrane nucleates Arp2/3 cross-linked actin networks with their barbed ends oriented toward the plasma membrane \cite{Blanchoin_PhysRev_2014}.
Our simple approach captures drop motion solely induced by active stress and polarization-surface (PS) coupling.
Surprisingly, motion does not depend on splay/bend configurations in the bulk liquid, but rather on the spatial variation of the polarization strength.
The onset of motion shows hysteresis, i.e., for a finite parameter range, resting and moving states are both stable. Strikingly, this bistability is reminiscent of the conversion between moving and resting keratocytes that can be triggered by mechanical or temperature stimuli \cite{VeSB1999cb,BKA2015pnasusa}. The hysteretic behavior persists for passive droplets without contractility. In this case, stationary asymmetric and symmetric drop shapes co-exist and correspond to different forms of internal organization of the drop's polar constituents. We provide a simple argument based on an effective free energy to account for the two competing forms of internal drop organization which might be relevant for the keratocyte system despite its biochemical complexity. A further transition at high active stresses may destabilize and split the drops.
\section{The model}
We consider a liquid drop composed of active polar constituents, which are oriented mainly parallel to the substrate.
For passive drops, the shape, spreading behavior and motion are crucially determined by wetting phenomena on the substrate \cite{DeGennes1985}, an effect which is neglected by most models for active drops. Here we employ a recently developed generic thin-film model that provides a framework for the dynamics of free-surface drops of partially wetting polar active liquids \cite{TSJT2020pre}. It naturally accounts for moving three-phase contact lines and allows for a systematic study of the role of nematic and polar order and their coupling to the free interface in inducing symmetry-breaking and motility.
The droplet behavior is described by two dynamic variables, the drop height profile $h(t,\mathbf{r})$ and the height-averaged polarization $\mathbf{p}(t,\mathbf{r})$, which is assumed to be parallel to the substrate.
\subsection{Underlying free energy of the passive polar droplet}
The free energy functional $\mathcal{F}$ defines the passive part of the evolution of $h$ and $\mathbf{p}$. Here we chose the minimal form
\begin{eqnarray}
\mathcal{F}[h,\mathbf{p}]   &=& \int  \Big[ \tfrac{ \gamma }{2} (\nabla h )^2 + f\lowt{w}(h)+
h f\lowt{spo}\left(h, \mathbf{p}^2\right) \nonumber  \\
& &+h f\lowt{el}(\nabla\mathbf{p}) 
+f\lowt{coupl}(\nabla h, \mathbf{p}) \Big] \mathrm{d}^2 r\,, \label{ap:en:hpspecific} 
\end{eqnarray}
where  $\nabla$ denotes the Nabla operator in two dimensions.
The first term in \eqref{ap:en:hpspecific}  represents the liquid-gas surface energy with the surface tension $\gamma$. For the wetting energy $f\lowt{w}$ we use a simple form for partially wetting liquids consisting of a long-range destabilizing van der Waals and a short-range stabilizing interaction analogous to Refs.~\cite{Pis2001pre,Thiele2010jpcm,TST+2018l}, i.e.,
\begin{equation}
f\lowt{w}=\frac{A}{2h^2}\left({2 h\lowt{a}^3\over 5h^3}-1\right)\,,
\end{equation}
where $A$ and $h\lowt{a}$ denote the Hamaker constant and the adsorption (or precursor) layer thickness, respectively. Note, that the equilibrium contact angle $\theta\lowt{eq}$ is related to $f\lowt{w}$ and $\gamma$ by $\cos(\theta_{\mathrm{eq}})=1+\frac{f_{\mathrm{w}}(h_{\mathrm{a}})}{\gamma}$, see e.g., \cite{StVe2009jpm}.
The function $f\lowt{spo}$ accounts for spontaneous polarization of the liquid, e.g., it may drive a transition between an isotropic, i.e., microscopically disordered state, and a polarized, i.e., microscopically ordered state. We employ the double-well energy
\begin{eqnarray}
f\lowt{spo}=-{c\lowt{sp}\over 2}\left[1-2\kappa(h)\right]\mathbf{p}\cdot\mathbf{p}+{c\lowt{sp}\over 4}(\mathbf{p}\cdot\mathbf{p})^2\,,\label{ap:fspo}
\end{eqnarray}
with $c\lowt{sp}>0$ and $\kappa(h)=(h\lowt{a}/h)^{6}$.
Through the height dependence $\kappa(h)$ (which is chosen to decay as the short-range interaction term in $f\lowt{w}(h)$), Eq.~\eqref{ap:fspo} allows for the existence of an ordered state $|\mathbf{p}|>0$ in the drop for $h\gg h\lowt{a}$ while the adsorption layer (where $h\approx h\lowt{a}$) remains in the disordered state $|\mathbf{p}|=0$ and does not influence the internal organisation of the drop.
The term $f\lowt{el}$ accounts for a liquid crystal elastic energy, with
\begin{equation}
f\lowt{el}={c\lowt{p}\over 2}\nabla\mathbf{p}:\nabla\mathbf{p}\,.
\end{equation}
The final contribution, $f\lowt{coupl}$, couples the polarization and the gradient of the free surface profile via the energy
\begin{equation}\label{eq:f_c}
f\lowt{coupl}={c\lowt{hpa}\over 2} \mathbf{p}\cdot\nabla h + {c\lowt{hpv}\over 2} \mathbf{p}\cdot\nabla_\bot h\,,
\end{equation}
where $\nabla_\bot h=(-\partial_{y} h, \partial_{x} h)\hight{T}$. We emphasize that $f\lowt{coupl}$ captures various types of polarization-surface (PS) coupling:
$c\lowt{hpa}>0$ favors an outward pointing aster-like polarization pattern, whereas $c\lowt{hpv}>0$ favors a vortex-like polarization pattern (counterclockwise).  
\subsection{Equations of motion of the active polar droplet}
The generalized evolution equations for the height profile $h$ and the components of the height-integrated polarization field $\mathbf{P}=h\mathbf{p}$ are given in their thermodynamic form by \cite{TSJT2020pre}
\begin{eqnarray}
\partial_t h&  = & \partial_\alpha \bigg[ Q_{hh}\Big(\partial_\alpha{\delta \mathcal{F}\over \delta h} -\mu_\alpha - \partial_\beta \sigma\hight{a}_{\alpha\beta} \Big) +Q_{hP_\beta}\partial_\alpha{\delta \mathcal{F}\over \delta P_\beta}  \bigg] \label{hP:ap:1}\\
\partial_tP_\gamma &= & \partial_\alpha \bigg[  Q_{h P_\gamma}\Big(\partial_\alpha{\delta \mathcal{F}\over  \delta h} -\mu_\alpha - \partial_\beta\sigma\hight{a}_{\alpha\beta}\Big) +  Q_{P_\gamma P_\beta}\partial_{\alpha}{\delta \mathcal{F}\over \delta P_\beta} \bigg]\notag\\
&&-Q\lowt{NC}{\delta \mathcal{F}\over \delta P_\gamma} \,.  \label{hP:ap:2}
\end{eqnarray}
\begin{figure*}[h]
\includegraphics[width=1.0\textwidth]{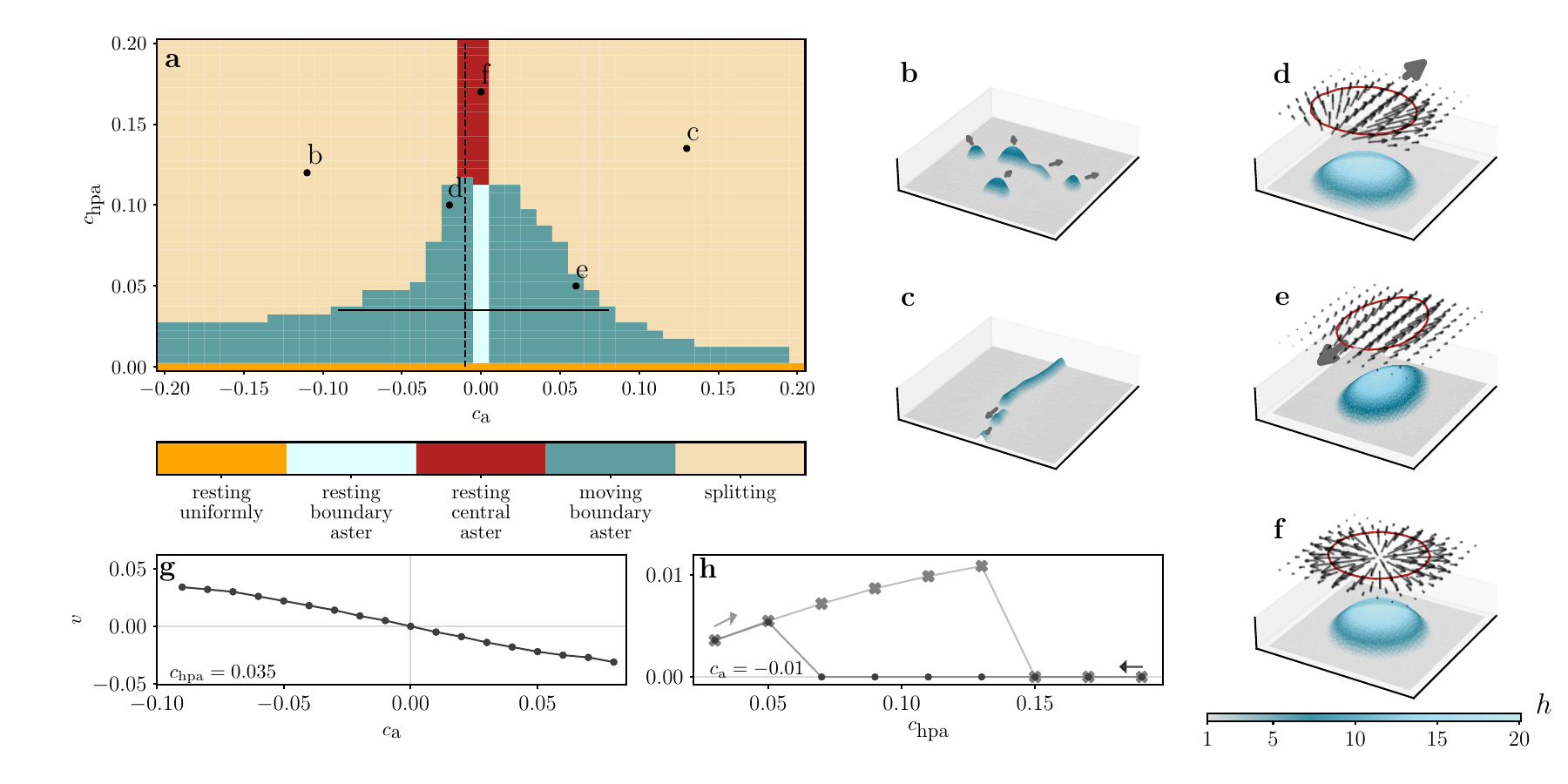}
\caption{Behaviour of active and passive 3D drops. (a) Morphological phase diagram in the plane spanned by the strengths of active stress $c\lowt{a}$ and PS coupling $c\lowt{hpa}$.
  (b-f) Snapshots of moving and resting drops [height profile (blue) with overlayed polarization (black) and contact lines (red), gray arrows indicate the direction of drop motion] at parameters indicated by symbols in (a). Drop velocity at constant $c\lowt{hpa}=0.035$ (g) and $c\lowt{a}=-0.01$ (h), as indicated by solid and dashed black lines in (a). In (h) hysteresis is shown by black dots [gray crosses] obtained when successively decreasing [increasing] $c\lowt{hpa}$, as indicated by the arrows. A positive [negative] velocity corresponds to motion parallel [antiparallel] to the drop's mean polarization. Shown domain sizes are $400\times400$ (b, c), and  $160\times160$ (d-f), which are only a small part of the full numerical domain $\Omega$.} \label{Fig:phase-diag}
\end{figure*}
It extends a gradient dynamics describing the overdamped dynamics of a mixture driven by the underlying energy functional \cite{TTL2013prl,XTQ2015JPCM} towards a polar liquid and combines this passive dynamics with active forces,
 namely, the active stress tensor $\boldsymbol{\sigma}\hight{a}=-c\lowt{a}\mathbf{p}\mathbf{p}$ (with $c\lowt{a}>0$ extensile and $c\lowt{a}<0$ contractile stress) and the self-propulsion force $\boldsymbol{\mu}=\mu_0{3\eta\over h^3}\mathbf{P}$. Here,
 $\eta$ and $\mu_0$ denote the dynamic viscosity of the fluid and the self-propulsion speed, respectively. $Q\lowt{NC}$ is a mobility ($\sim$ rotational diffusivity) related to the non-conserved flux in the polarization field.
The remaining mobilities in Eqs.~\eqref{hP:ap:1} and \eqref{hP:ap:2} are given by
\begin{eqnarray}
Q_{hh} &= & {h^3\over 3\eta}\\
Q_{hP_\alpha} & = & {h^2P_\alpha\over 3 \eta}\\
Q_{P_\alpha P_\beta} & = & h\left({P_\alpha P_\beta \over 3\eta}+M\delta_{\alpha\beta}\right)\, ,\label{eq:mob3}
\end{eqnarray}
where $M$ is the self-diffusivity of the polarization field. An extensive discussion of the form \eqref{hP:ap:1}-\eqref{eq:mob3} is found in Ref.~\cite{TSJT2020pre}.
Finally, for comparison with the literature we write Eqs.~\eqref{hP:ap:1} and \eqref{hP:ap:2} in hydrodynamic form
\begin{eqnarray}
\partial_th & = & -\nabla\cdot\left(\mathbf{j}^\mathrm{Cv} +\mathbf{j}^\mathrm{Ca}\right)\\
\partial_tP_\alpha& =& -\nabla\cdot\left[p_\alpha \left(\mathbf{j}^\mathrm{Cv}+\mathbf{j}^\mathrm{Ca}\right)+\mathbf{j}^{\mathrm{D}P_\alpha}\right]+j^{\mathrm{R}}_\alpha
\end{eqnarray}
with the passive fluxes 
\begin{eqnarray}
\mathbf{j}\hight{Cv} & =& -{h^3\over 3\eta}\left[\nabla{\delta \mathcal{F}\over \delta h}+{P_\beta\over h}\nabla{\delta \mathcal{F}\over \delta P_\beta}\right],\\
\mathbf{j}^{\mathrm{D}P_\alpha} &= &-hM\nabla{\delta \mathcal{F}\over \delta P_\alpha},\\
j^{\mathrm{R}}_\alpha & = & -Q\lowt{NC} {\delta \mathcal{F}\over \delta P_\alpha}\,.
\end{eqnarray}
and the active flux
\begin{equation}
\mathbf{j}^\mathrm{Ca} = {h^3\over 3\eta}\nabla\cdot\boldsymbol{\sigma}\hight{a}+\mu_0\mathbf{P}\,.\label{jca}
\end{equation}
Note, that the above introduced framework uses a similar minimal set of elastic, nematic and polar ingredients as  Ref.~\cite{TTMC2015nc} with the important difference, that here we employ a thin-film approximation including a wetting energy, whereas Ref.\ \cite{TTMC2015nc} uses a 3D phase-field approach, where the contact angle between the ``eukaryotic cell'' and the substrate is fixed at $\pi/2$. In contrast, in our model the contact angle is fully dynamic and naturally evolves with the shape of a drop moving through activity.
In the following, we neglect self-propulsion, i.e., we set $\mu_0=0$, since with self-propulsion the resulting emergence of droplet motion is evident. Instead, here we concentrate on the competition between elastic energy in the bulk ($f\lowt{el}$), PS coupling ($f\lowt{coupl}$) and the strength of active stresses (mainly contractile $c\lowt{a}<0$). In particular, we focus on aster-like structures resulting from PS coupling mechanisms ($c\lowt{hpa}>0$, $c\lowt{hpv}=0$), reminiscent of the organization of actin filaments (barbed ends pointing towards the cell membrane) in living cells.
\section{Phase diagram of active drops: Motility and drop splitting}
\begin{figure}[h]
\includegraphics[width=0.5\textwidth]{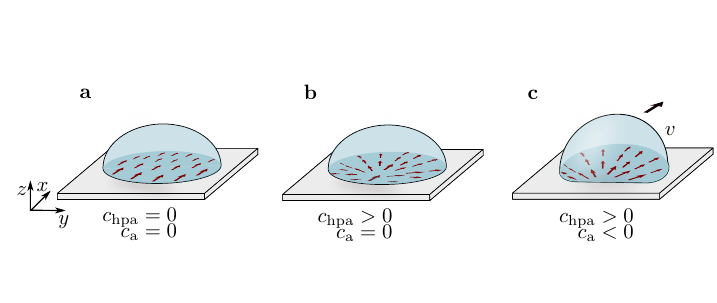}
\caption{Sketch of passive and active drops on a solid substrate. The height-averaged polarization is shown as red arrows. The strengths of active stress $c\lowt{a}$ and the PS coupling $c\lowt{hpa}$ allow one to distinguish:
(a) uniform polarization in a passive drop without PS coupling;
(b) boundary aster in a passive drop with weak PS coupling;
(c) moving active drop of back-forth asymmetric shape for contractility and PS coupling. The direction of motion is indicated by the black arrow.
}
\label{Fig:sketch}
\end{figure}

%
Figure \ref{Fig:phase-diag}(a) presents a morphological phase diagram obtained by direct time simulations as described in the appendix. It summarizes the role of active stress and PS coupling for drop motility and morphology for a constant imposed elastic stress.  Without PS coupling [($c\lowt{hpa}=0$, orange region at bottom of Fig.~\ref{Fig:phase-diag}(a)], drops polarize uniformly, avoiding any defect structures in the bulk. These \textit{uniformly polarized} drops remain at rest not only in the passive case but also at any active stress since they do not develop a back-forth asymmetry with respect to the (nematic) active stress. Nevertheless, the drop shape adapts to the active stress-type.
Increasing the PS coupling, the drop behavior becomes more complex. Passive drops ($c\lowt{a}=0$) are necessarily always at rest but change their polarization state: From uniformly polarized, to a \textit{boundary aster} state with a half aster-like defect at the contact line [light blue region in Fig.~\ref{Fig:phase-diag}(a)] to a completely radially symmetric \textit{central aster} state with outward pointing polarization [red region in Fig.~\ref{Fig:phase-diag}(a) and example profile in Fig.~\ref{Fig:phase-diag}(f)]. The transition results from an increasing dominance of the energy gain due to PS coupling over the elastic energy cost due to a nematic defect in the bulk, i.e., $f\lowt{coupl}$ dominates $f\lowt{el}$. 
The described sequence of structural changes is also essential in the active case ($c\lowt{a}\neq 0$): At weak PS coupling, the back-forth symmetry in drop shape and polarization is broken resulting in steadily moving drops [dark blue region in Fig.~\ref{Fig:phase-diag}(a) and profiles in Fig.~\ref{Fig:phase-diag}(d) and (e)].  With active contractility the drop contracts along the direction of its net polarization and the back-forth asymmetry in the polarization translates into steady drop motion. The defect is located at the trailing edge of the drop [see \ref{Fig:phase-diag}(d)], consistent with results of fully 3D hydrodynamic models \cite{TTMC2015nc} and the planar case \cite{MWP2015jotrsi,TMC2012potnaos}.
In contrast, an active extensile stress elongates the drop in the direction of its net polarization. Transiently, liquid is locally accumulated near the defect before the energetic cost of surface and elastic energy reorients the polarization field and forms a steadily moving drop as shown in Fig.~\ref{Fig:phase-diag}(e). Here the advancing drop edge coincides with the location of the defect.  
For contractile active stress the above described mechanism of structural polarity and onset of motion is illustrated in Fig.~\ref{Fig:sketch}.
At strong PS coupling, a large active stress overcomes the stabilizing effect of surface tension and large drops split into smaller ones [yellow region in Fig.~\ref{Fig:phase-diag}(a)].
Subsequently, splitting continues until surface tension dominates. The result are ensembles of smaller steadily moving drops [see snapshots in Fig.~\ref{Fig:phase-diag}~(b) and (c)]. A phase diagram for smaller droplets is shown and discussed in the appendix (Fig.~\ref{Fig:phase-diag-app}).  Then, for instance, the transition to droplet splitting occurs at larger active stress because the destabilizing active bulk flow is weaker than the stabilizing surface tension that becomes more dominant for smaller drops.
Fig.~\ref{Fig:phase-diag}(g) shows how drop speed  monotonically increases with active stress $c\lowt{a}$ before drops split at strong PS coupling (where the black line ends). Strikingly, a close inspection in the range of low contractile stress reveals that hysteresis between moving and resting drops occurs when changing the strength of PS coupling $c\lowt{hpa}$ [Fig.~\ref{Fig:phase-diag}(h)]. Starting with an axisymmetric resting drop at $c\lowt{hpa}=0.19$, a successive decrease in $c\lowt{hpa}$ (black dots) results in the transition to a moving polar drop at $c\lowt{hpa}\approx0.07$. However, starting with a moving drop at small $c\lowt{hpa}=0.03$, a successive increase in $c\lowt{hpa}$  (gray crosses) only stops the drop at $c\lowt{hpa}\approx 0.15$. Such a coexistence of stable resting and stable moving states is experimentally well documented for keratocytes \cite{VeSB1999cb,BKA2015pnasusa}, but a conclusive mechanistic picture has been missing.
A deeper understanding of the underlying mechanism of motion and bistability can be gained by considering simpler transversally invariant liquid ridges (2D drops). By employing continuation techniques \cite{DWCD2014ccp, UeWR2014nmma}, we can follow stable \textit{and} unstable droplet states over a large range of contractility at low computational cost. This provides clear information on hysteretic behavior.
However, if splay and bend instabilities in the polarization field were necessary to drive drop motion with active stresses as abundantly argued in \cite{MWP2015jotrsi,TMC2012potnaos,TTMC2015nc}, no motion should be found for liquid ridges with a polarization mainly parallel to the 1D substrate, i.e., without splay or bend. In contrast, our computations for 2D drops reveal qualitatively identical behavior as described above for 3D drops, and enable us to further scrutinize the transitions between resting and moving drops in ridge geometries without losing crucial motility features.
\begin{figure*}[h]
\includegraphics[width=1.0\textwidth]{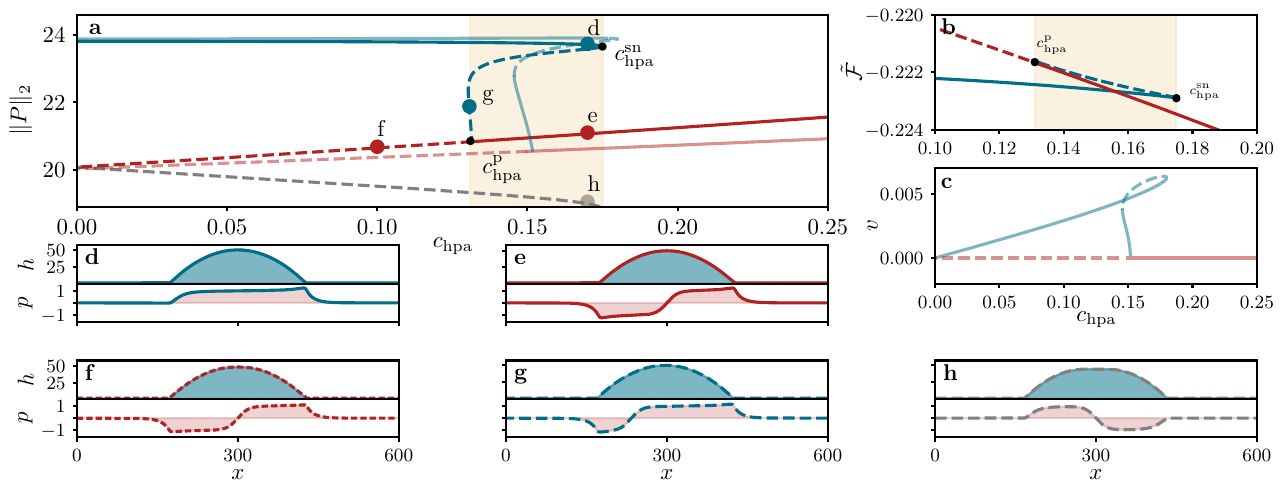}
\caption{
Hysteresis and motility in 2D drops (ridges). (a) Bifurcation diagram using the L$_2$ norm of $P$ over the strength of PS coupling $c\lowt{hpa}$ for drops with a boundary defect (blue) and with a central defect (red and gray). Strong colors indicate the passive case $c\lowt{a}=0$, light colors indicate the active contractile case $c\lowt{a}=-0.005$.
The pitchfork [saddle-node] bifurcation is marked $c\lowt{hpa}\hight{p}$ [$c\lowt{hpa}\hight{sn}$]. Dashed [solid] lines correspond to unstable [stable] states and the bistability region is shaded yellow.
(b) Free energy for central [boundary] defect states in red [blue] for passive drops. (c) Velocity of active drops ($c\lowt{a}=-0.005$) for central [boundary] defect states in light red [blue]. (d-h) Height (top) and polarization (bottom) profiles at $c\lowt{hpa}$ values indicated in (a).}\label{Fig:bif-chp}
\end{figure*}
\section{Structural hysteresis in passive and active 2D drops}
%
Fig.~\ref{Fig:bif-chp}(a) shows the bifurcation behavior of passive ridges ($c\lowt{a}=0$, lines in strong colors) in dependence of the PS coupling strength $c\lowt{hpa}$. Corresponding drop and polarization profiles are shown in Figs.~\ref{Fig:bif-chp}(d-h).
At small $c\lowt{hpa}$ stable uniformly polarized drops with a back and forth asymmetry [solid blue curve, e.g., Fig.~\ref{Fig:bif-chp}(d)] coexist with unstable drops containing a central polarization defect [dashed red and gray curves, e.g., Fig.~\ref{Fig:bif-chp}(f,g)]. At high $c\lowt{hpa}$ only (stable) symmetric drops with outward pointing polarization and a central defect exist [solid red dashed line, e.g., Fig.~\ref{Fig:bif-chp}(e)]. The corresponding polarization profile in 2D ridges is equivalent to the axisymmetric aster structure for the 3D drops. The transition between the weak and strong PS coupling is marked by a complex bifurcation behavior with a multistable parameter range.
Unstable drops with a central polarization defect gain stability in a pitchfork bifurcation at $c\lowt{hpa}\hight{p}=0.131$. The emerging branch bifurcates subcritically and is unstable [dashed blue curve e.g., Fig.~\ref{Fig:bif-chp}(g)]. It represents boundary defect states. Following this branch, we reach a saddle-node bifurcation at $c\lowt{hpa}\hight{sn}=0.175$ and observe that the defect is further shifted toward the drop edge until it finally coincides with it. At the saddle-node bifurcation the branch turns back and becomes stable [solid blue line].
Remarkably, the transition between the two stable states (central defect vs. boundary defect) shows a hysteresis between $c\lowt{hpa}\hight{p}$ and  $c\lowt{hpa}\hight{sn}$ [yellow region in Fig.~\ref{Fig:bif-chp}(a)]. This is also illustrated by the representation of the free energy (fixed drop volume)  $\tilde{\mathcal{F}}$ in Fig.~\ref{Fig:bif-chp}(b): At small $c\lowt{hpa}$, the boundary defect state corresponds to the global energy minimum, while at large $c\lowt{hpa}$ the central defect is as the only persisting state the global minimum. Within the yellow hysteresis region, both states correspond to energy minima and are separated by a saddle point in the energy landscape.
\begin{figure*}[h]
\includegraphics[width=1.0\textwidth]{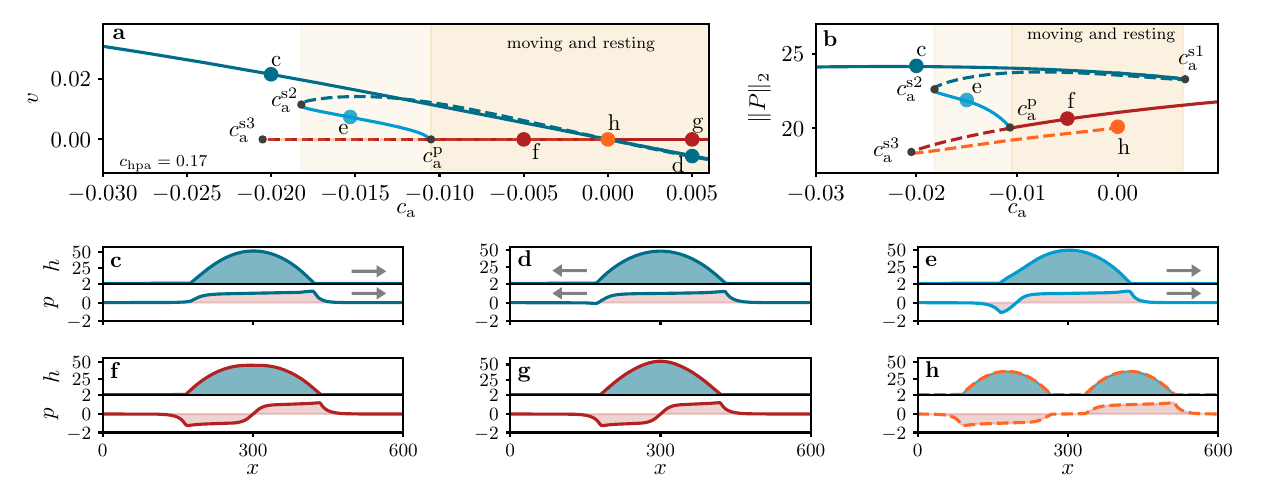}
\caption{
Bifurcation diagram for active polar 2D drops (ridges). (a) Velocities $v$ of drops with a boundary defect (dark and bright blue) and a central defect (red) are given as a function of active stress $c\lowt{a}$ at fixed $c\lowt{hpa}=0.17$. (b) Bifurcation diagram related to the hysteretic transition between resting and moving drops and drop splitting. Shown is the L$_{2}$ norm of $P$ over $c\lowt{a}$. Labels $c\lowt{a}\hight{s1}$ to $c\lowt{a}\hight{s3}$, $c\lowt{a}\hight{p}$ are explained in the main text. Dashed [solid] lines correspond to unstable [stable] states. The strongly [weakly] yellow-shaded region indicates the bistability regime of moving and resting [two different moving] droplets. (c-h) Film height (top) and polarization profiles (bottom) at $c\lowt{a}$ values as indicated by symbols in (a,b).}\label{Fig:ca}
\end{figure*}
\subsection{Bistability between motile and stationary states in active 2D drops}
The structural organization of the analyzed passive drops is maintained in the active case, however, all states with back and forth asymmetry, i.e., with boundary defects are motile. A corresponding bifurcation diagram for active contractile drops ($c\lowt{a}=-0.005$) is shown in light colors in Fig.~\ref{Fig:bif-chp}(a) while the corresponding drop velocities are shown in Fig.~\ref{Fig:bif-chp}(c). 
With increasing contractility, the pitchfork bifurcation at $c\lowt{hpa}\hight{p}$ is shifted towards larger values of $c\lowt{hpa}$. Interestingly, an additional saddle-node bifurcation appears on the blue branch rendering the pitchfork bifurcation at $c\lowt{hpa}\hight{p}$ supercritical. In consequence,  an additional sub-branch of stable moving drops exists in the active case. Hence, there is now a range of bistability between two different moving states adjacent to the range of bistability between a moving and a resting state.
To further clarify the active droplet behavior we also studied the influence of the active stress at constant PS coupling ($c\lowt{hpa}=0.17$ where hysteresis occurs). A corresponding bifurcation diagram is presented in Fig.~\ref{Fig:ca} where panels (a) and (b) show the drop velocity and L$_2$ norm of $P$, respectively. Exemplary drop and polarization profiles are shown in panels (c-h). Briefly, for contractile stress we find a hysteresis regime with motile droplets coexisting with stationary droplets [dark yellow region in Fig.~\ref{Fig:ca}~(a,b)]. In addition, there exists a complex bifurcation scenario related to drop splitting. An important observation is that back and forth asymmetric stable drops with a boundary defect are motile at arbitrarily low stress.
In addition, a second stable motile asymmetric \textit{off-center defect} drop [light blue solid branch in Fig.~\ref{Fig:ca}(a)] emerges at a finite contractile stress via a pitchfork bifurcation (at $c\lowt{a}\hight{p}=-0.0105$) that renders the stationary symmetric drops with a central defect [red branch in Fig.~\ref{Fig:ca}(a)] unstable. The two coexisting types of motile drop move at different speeds. They are connected via an unstable branch and two saddle-node bifurcations (at $c\lowt{a}\hight{s1}=0.0068$ and $c\lowt{a}\hight{s2}=-0.0177$). The direction of droplet motion depends on the type of active stress and the direction of net polarization, i.e., right-polarized drops move to the right [left] for contractile [elongational] stress [Fig.~\ref{Fig:ca}(c,d)].
Stationary drops with a central defect only exist for $c\lowt{a}>c\lowt{a}\hight{s3}=-0.0207$ where they merge via a saddle-node bifurcation with a branch of split drops [orange dashed in Fig.~\ref{Fig:ca}(a,b)]. The latter is unstable and ends at $c\lowt{a}=0$ on a branch related to the coarsening of passive drops (not shown). A more extensive representation of the behavior of 2D droplets at different PS couplings $c\lowt{hpa}$ and varying active stress $c\lowt{a}$ can be found in the morphological phase diagram in Fig.~\ref{Fig:pd-app} in the appendix. The results of Figs.~\ref{Fig:bif-chp},~\ref{Fig:ca} correspond to cuts through this diagram.
\subsection{A passive structural hysteresis at the origin of bistability in cell crawling}
\label{sec:toy}
\begin{figure}[htbp]
\includegraphics[width=0.5\textwidth]{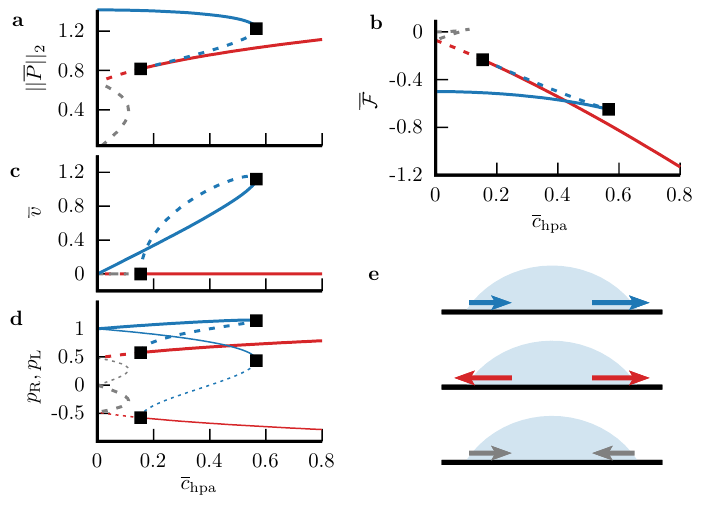}
\caption{PS coupling at the origin of a structural transition with hysteresis in polar droplets in a toy model [Eq.~\eqref{eq:simo}]. Bifurcation diagrams giving (a) the L$_2$ norm of the polarization, (b) free energy, (c) the droplet speed $\bar v=p\lowt{R}^2-p\lowt{L}^2$, and (d) the polarizations $p\lowt{R}$ and $p\lowt{L}$ in the presence of active stress using the PS coupling strength $\bar{c}\lowt{hpa}$ as control parameter.  The line colors correspond to the three different polar organizations shown schematically in (e). Thick and thin lines in (d) correspond to $p\lowt{R}$ and $p\lowt{L}$, respectively. The black square symbols delimit the range of coexistence of stable polar and stable symmetric drops (solid blue and solid red lines). The elastic constant is $\bar{c}\lowt{p}=0.2$. \label{Fig:sm}}
\end{figure}

The occurrence of moving ridges clearly evidences that liquid motion results from a spatial modulation of the strength and not the orientation of polarization. Although splay may contribute to motion in 3D drops, the described moving ridges directly point to the polarization strength as primary origin of motility. The similarity of velocities in our 2D and 3D cases [Figs.~\ref{Fig:phase-diag}(g) and \ref{Fig:ca}(a)] suggests that splay is of little importance.
A central feature of the complex bifurcation diagrams in Figs.~\ref{Fig:ca} and \ref{Fig:bif-chp} are the regions of bistability of a polar moving state and a symmetric resting state. This bistability qualitatively corresponds to the hysteresis behavior for the transition of a resting aster state to a moving boundary aster state in 3D.
In the biological context, this means that at a given contractile stress $c\lowt{a}$ and PS coupling $c\lowt{hpa}$ drops (cells) may coexist in a stable moving state with uniform polarization and a stable symmetric resting state. The two states are separated by an unstable (moving) threshold state, i.e., the interconversion between resting and moving states requires a finite perturbation. For example, an insufficiently strong perturbation of the resting state only results in transient motion. Increasing the contractile stress beyond a critical value, the resting state loses stability and a persistently moving state is reached. 
These qualitative features may be illustrated employing a toy model based on a simple energy argument neglecting the complex droplet dynamics. It emphasizes the key idea that orientational order induced by the droplet boundaries and the energetic costs of defects in the bulk facilitate a transition between symmetric and polar droplet states.
We consider a simplified 1D droplet, whose polarity is described by two variables, i.e., the polarizations at the right and left boundary, $p\lowt{R}$ and $p\lowt{L}$, respectively. Neglecting spatial gradients and contributions from the free interface we can write 
the free energy $\bar{F}$ in terms of $p\lowt{R}$ and $p\lowt{L}$
\begin{eqnarray}
\bar{\mathcal{F}} & =& {1\over 4}\left(p\lowt{R}^4+p\lowt{L}^4\right)- {1\over 2}\left(p\lowt{R}^2+p\lowt{L}^2\right)+\nonumber\\
 & & {\bar{c}\lowt{p} \over 4} \left(p\lowt{R}-p\lowt{L}\right)^4+{\bar{c}\lowt{p} \over 2} \left(p\lowt{R}-p\lowt{L}\right)^2- \label{eq:simo}\\
 & & \bar{c}\lowt{hpa}\left(p\lowt{R}-p\lowt{L}\right)\nonumber\,. 
\end{eqnarray}
The first line in  Eq.~\eqref{eq:simo} describes the energy of polarization at the right and left interfaces as a  respective double-well potential that allows for two polarization directions, the second line encodes the elastic energy due to the difference in the polarization directions at the two interfaces and the last line introduces the PS coupling. In the presence of an active contractile nematic stress $\sim \mathbf{pp}$, a difference in the magnitude of polarization $p\lowt{R}^2\neq p\lowt{L}^2$ will induce a pressure difference between the two interfaces and hence droplets experience a driving force and move with a speed $\bar v\sim p\lowt{R}^2-p\lowt{L}^2$.  
The passive structural organization of such simple droplets [Fig.~\ref{Fig:sm}(d)] and the resulting bifurcation diagrams [Fig.~\ref{Fig:sm}(a,b)] are all given by the extrema of the potential \eqref{eq:simo}. A comparison of Figs.~\ref{Fig:bif-chp} and \ref{Fig:sm} shows that the results obtained with this simple scheme resemble in its essential features the behavior of the full dynamical model. Namely, there is a polar state which moves in the presence of active stress, and two symmetric states (one stable and one unstable) that are at rest. Even the structural hysteresis is present as stable stationary symmetric and stable polar (moving) states share a range of coexistence in the PS coupling strength. 

\section{Conclusions}
To conclude, we have modeled shallow three-dimensional (3D) active free-surface drops with a mean polar order parallel to the substrate. Incorporating a polarization-surface (PS) coupling has allowed us to identify a mechanism of motility in which active stresses drive drop motion through spatial variations of polarization strength. Motility is induced even for liquid ridges (2D drops), implying that splay in the polarization field is not essential for the mechanism. This provides an alternative to previous proposals based on splay or bend \cite{TMC2012potnaos,TTMC2015nc,MWP2015jotrsi}. 
Living cells exert a tight control over the cytoskeleton via signaling pathways which are partially localized in the cell membrane, i.e., the machinery necessary for the growth and architecture of polar Arp2/3 cross-linked actin networks and linear filopodia structures induced by formins is localized at the plasma membrane \cite{Blanchoin_PhysRev_2014}. Within the context of active drops as minimal models for living cells a fine tuning of the PS coupling by the cell is a reasonable assumption. 
Thus,  the physico-chemical and mechanical properties of active polar filament assemblies may  play an important role in the robustness of cell migration in addition to complex biochemical signaling pathways.\\
Furthermore, our model provides a possible mechanism of bistability in the keratocyte system based on well-identified physical components (encoded in a variational model) and active stress, without the need for complex biochemical feedback mechanisms \cite{BKA2015pnasusa,RMG2020PRR,ZiSA2012jrsi}.
\section*{Acknowledgements}
We acknowledge the doctoral school ``Active living fluids'' of the German French University (Grant No. CDFA-01-14), and the DAAD \& Campus France (PROCOPE 2020 Grant No. 57511756). FS thanks the ``Studienstiftung des deutschen Volkes'' for financial support.
\section*{Author Contributions}
F.S. performed the simulations for Figs.~\ref{Fig:phase-diag} to \ref{Fig:ca} and K.J. performed simulations for Fig.~\ref{Fig:sm}. All authors together developed the interpretation and progression of modeling and wrote the manuscript.
\section*{Conflicts of interest}
There are no conflicts to declare.
\section*{Appendix A}
\subsection*{Numerical details}
Direct time simulations are performed employing the open source library \texttt{oomph-lib} \cite{HeHa2006}, which solves the evolution equations using Finite Elements and Newton's method. For the time steps Backward Differentiation Formulas  of second order (BDF2) are employed. An adaptive grid is used for spatial resolution, in the case of 3D drops for domain size $\Omega=800\times800$ and for 2D drops for domain size $L=600$, in each case with periodic boundary conditions in $x$-direction (and in the 3D case additional no-flux boundary conditions in $y$-direction). All 3D simulations were started with a paraboloid of height $h_{0}=20$ with drop volume $V_{3\mathrm{D}}=\int_{\Omega}h\,d\mathbf{x}=71822$ above precursor $h\lowt{a}$, and all 2D simulations (including path-continuation) correspond to drop volume of $V_{2\mathrm{D}}=\int_{0}^{L}h\,d x=8280$ above $h\lowt{a}$.
Simulations for Fig.~1 were initiated with small, random polarization into $x$ direction and $c\lowt{a}$ was introduced into the system at $t=5\times 10^{4}$ with $T=10^{5}$ the overall time of simulations. For moving and resting states some simulations are continued beyond the usual $t=10^{5}$ up to $t=10^{12}$ increasing the likelihood that these states are steadily moving and not only transients. 
In Fig.~1(b-f), red lines indicate the contact lines of the drops' shape. They are obtained as contour lines of the height field slightly above the adsorption layer thickness, namely, at $h=3.0$ in Fig.~1(d-f), at $h=1.5$ in Fig.~1(b,c).
For the purpose of continuation, the equations of motion (6) and (7) are transformed into a co-moving reference frame with velocity $v$. Based on the pseudo-arclength continuation, resting and moving states can be followed in parameter space (using the velocity $v$ as additional degree of freedom) employing \texttt{pde2path} \cite{UeWR2014nmma} or \texttt{auto07p} \cite{DWCD2014ccp}. The bifurcation diagrams in Figs.~3 and 4 were created using the homotopy method, i.e., by successively changing various parameters from zero to increase the degree of nonlinearity.
\subsection*{Parameter settings}
If not stated otherwise the following parameters are fixed for all calculations: $c\lowt{p}=2.0$, $c\lowt{sp}=0.01$, $Q\lowt{NC}=h\lowt{a}=M=A=\eta=1.0$, $\alpha_{0}=0$.\\
Decreasing $Q\lowt{NC}$ changes the bifurcation diagram as the dynamics of the polarization field is increasingly dominated by the conserved, i.e., convective, contribution. The parameter thus affects the occurrence of bistability between boundary and central defect states in the passive case, so that it cannot be found for certain values of $Q\lowt{NC}$. Thus, the additional specification of contractility would mean that no bistability can be found between stationary and moving states. In a biophysical context, however, it seems highly unlikely that the dynamics of the polarization is nearly conserved. On the contrary, the polarization should be able to change rapidly, which legitimizes the choice of $Q\lowt{NC}=1$.
Note, that bistability is very robust when varying the parameter $M$, i.e., the bifurcation diagram in Fig.~3 does (merely) not change. Other parameter choices result from nondimensionalization, thus are not arbitrary. We set $Q\lowt{NC}=1$ and $M=1$ to ensure that all occurring processes happen on similar time scales.
\subsection*{Phase diagram for droplets of small volume}
\begin{figure}[htbp]
\centering
\includegraphics[width=1\hsize]{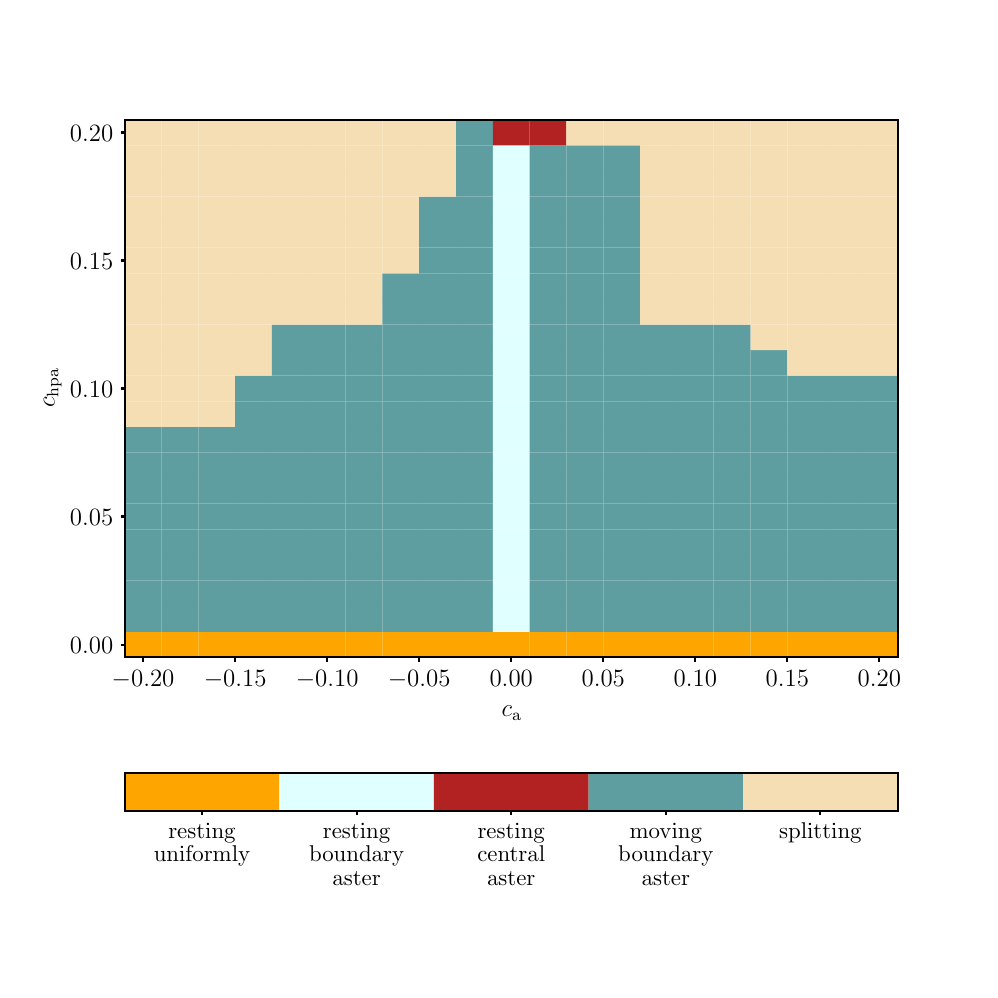}
\caption{Morphological phase diagram for 3D droplets as in Fig.~\ref{Fig:phase-diag}(a), however, for droplets with approximately a fifth of the volume. The occurring states are as in Fig.~\ref{Fig:phase-diag}(a), however, the transitions take place at different values of $c_{\mathrm{a}}$ and  $c_{\mathrm{hpa}}$.} \label{Fig:phase-diag-app} 
\end{figure}

 We have additionally performed a parameter scan comparable to Fig.~\ref{Fig:phase-diag}(a) in the plane spanned by $c_\mathrm{a}$ and $c_\mathrm{hpa}$, however, this time for droplets which have a considerably smaller volume than the ones in Fig.~\ref{Fig:phase-diag}(a). The resulting phase diagram is given in Fig.~\ref{Fig:phase-diag-app}. Although it qualitatively resembles Fig.~\ref{Fig:phase-diag}(a), the boundary where moving or resting drops become unstable and drop splitting sets in is shifted to higher active stresses. This highlights the stabilizing role of surface tension.  As surface forces play an increasingly important role with decreasing drop size larger active bulk forces are needed for droplet splitting to occur.
Moreover, the structural transition for passive systems ($c_\mathrm{a}=0$, from boundary aster (light blue in Fig.~\ref{Fig:phase-diag-app}) to central aster (red) is shifted to stronger PS coupling. Here the smaller droplet size heavily penalizes the formation of the central defect (i.e., the creation of a defect is more unfavorable compared to the bulk polarization energy in small drops than in large drops), requiring a stronger PS coupling.
\subsection*{Phase diagram for 2D drops (ridges)}
\begin{figure}[htbp]
\includegraphics[width=0.5\textwidth]{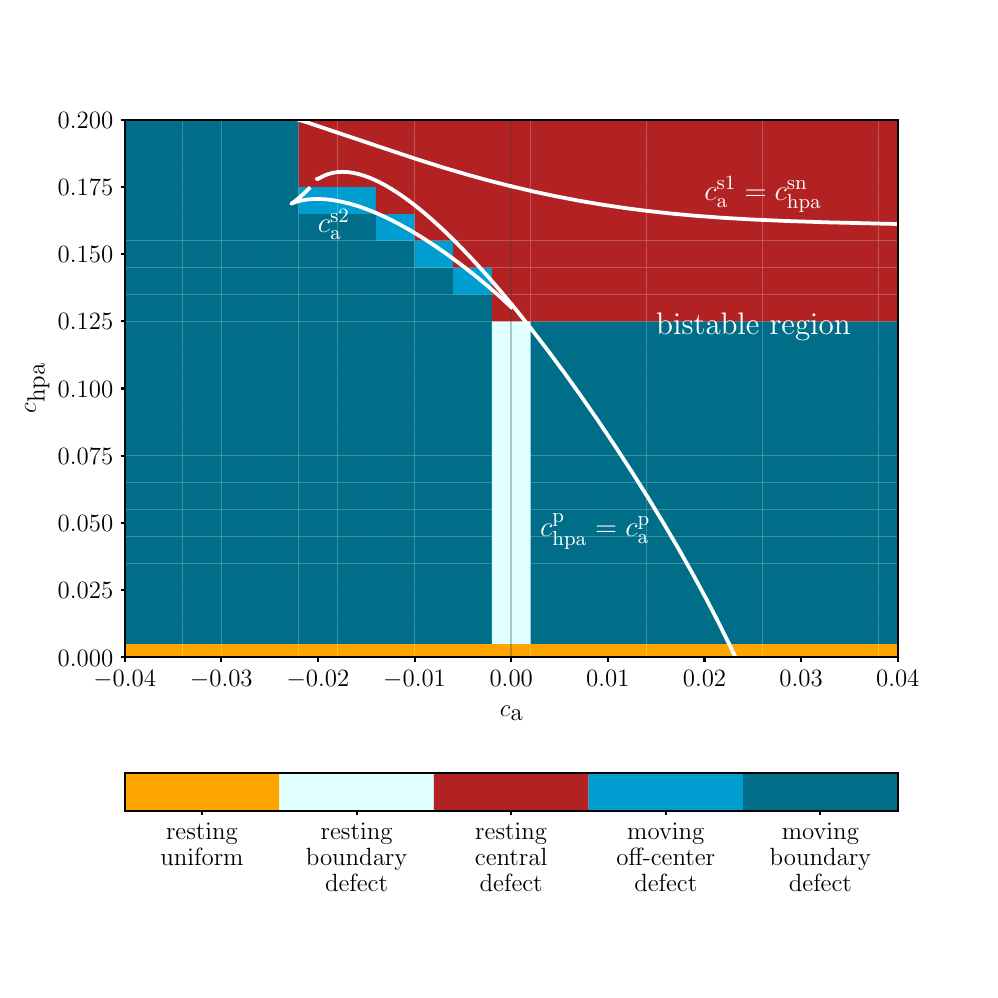}
\caption{Phase diagram for 2D droplets obtained by direct time simulations and also including loci of saddle-node and pitchfork bifurcations obtained by continuation. Each time simulation is initiated with a passive droplet at the respective PS-coupling $c_\mathrm{hpa}$ and after an equilibration period (more details in the text) an additional active stress $c_\mathrm{a}$ is introduced. The final states are characterized according to the color bar given at the bottom. The region where bistability between resting and moving droplets occurs lies between the two white lines $c_\mathrm{a}^\mathrm{s1}$ and $c_\mathrm{hpa}^\mathrm{p}$, i.e., the respective saddle-node and pitchfork bifurcations. Another bistable region exists between the two white lines marked $c_\mathrm{a}^\mathrm{p}$ and $c_\mathrm{a}^\mathrm{s2}$. There, moving states with a boundary defect and with an off-center defect both exist as stable states. The bifurcation lines are only shown where they are relevant for regions of bistability.}\label{Fig:pd-app}
\end{figure}

In the main text two bifurcation diagrams [cf.~Figs.~\ref{Fig:bif-chp} and \ref{Fig:ca}] are presented showing the relation between various branches of passive and active 2D droplets, i.e., ridges. The interplay of the various bifurcations explains the origin of parameter regions of bistability. This is further illustrated in the morphological phase diagram in Fig.~\ref{Fig:pd-app}, again spanned by the active stress $c_\mathrm{a}$ and the PS coupling $c_\mathrm{hpa}$.  
It is compiled on the one hand by direct time simulations, performed in a similar way as described above for the case of 3D droplets in Fig.~\ref{Fig:phase-diag}.
On the other hand, we directly focus on the onset of motion and the most interesting bistable region. To do so, we use continuation methods to directly track saddle-node and pitchfork bifurcations in parameter space. This has allowed us to identify the region of coexistence between moving and resting droplets (labeled ``bistable region'' in Fig.~\ref{Fig:pd-app}). In addition, we have identified the coexistence region of two different moving states, which we show and discuss already in Figs.~\ref{Fig:bif-chp} and \ref{Fig:ca}.
{\sl Parameter Coninuation:} We start the continuation from the passive case, and follow the pitchfork bifurcation  $c\lowt{hpa}\hight{p}$ in Fig.~\ref{Fig:bif-chp} when $c_\mathrm{a}$ is varied. This bifurcation actually is identical to the pitchfork bifurcation at $c\lowt{a}\hight{p}$ in Fig.~\ref{Fig:ca}~(a,b).  This becomes clear when looking in Fig.~\ref{Fig:pd-app} at a cut along the horizontal defined by $c\lowt{hpa}=0.17$. Furthermore, we track the saddle-node bifurcation at $c\lowt{hpa}\hight{sn}$ in Fig.~\ref{Fig:bif-chp} when changing $c\lowt{a}$ which is identified with the saddle-node bifurcation at $c\lowt{a}\hight{s1}$ in Fig.~\ref{Fig:ca}~(a,b). The parameter range between these two bifurcation corresponds to the region of bistability.
{\sl Direct Simulations:} The outcome of the direct time simulations heavily depends on the pathway taken into the stable state, which is why we briefly describe the numerical development in the following.
The simulations start with a small random polarization, which first grows at the drop edges and then spreads into the bulk of the drop. If the PS coupling is zero, in this way a uniformly polarized droplet is formed. However, as soon as the coupling deviates from zero, i.e., $c\lowt{hpa}>0$, the energy contribution of the PS coupling initially predominates and opposite polarizations form at the two drop edges [cf.~Eq.~\eqref{eq:f_c}]. These develop initially into a central defect over time. 
Depending on the strength of the PS coupling [cf.\ energy of the central vs boundary defect states in Fig.~\ref{Fig:bif-chp}(b)], the polarization may further evolve into a boundary defect due to the prevailing elastic energy contribution $f\lowt{el}$. However, if the coupling lies in the bistable region [cf.\ yellow region in Fig.~\ref{Fig:bif-chp}~(b)], the central defect state corresponds to a local minimum in the free energy. Therefore, it remains linearly stable and it takes a strong perturbation to shift the state to the global minimum, i.e., the boundary defect state [corresponding to the blue line in Fig.~\ref{Fig:bif-chp}~(b)]. The transition from the boundary defect to the central defect in the performed direct time simulations, thus, occurs at $c\lowt{hpa}=0.13$ [$=c\lowt{hpa}\hight{p}$ in Fig.~\ref{Fig:bif-chp}(b)] and the global minimum of the free energy is not always adopted.
Since all the time simulations identically start with a passive drop (at respective $c\lowt{hpa}$ values) before active stress is added, the sharp horizontal border between the blue and red areas at $c\lowt{a}\ge0$ directly reflects the occuring bistability in the passive case, i.e., along the vertical line at $c\lowt{a}=0$.
For values of active stress greater than $c\lowt{a}\hight{s1}$, only the stable resting central defect states exist, which is confirmed by the direct time simulations. For values smaller than $c\lowt{a}\hight{p}$, only moving states with off-center and boundary defect are stable. In the region between $c\lowt{a}\hight{p}$ and $c\lowt{a}\hight{s2}$ we mainly find the off-center defect state where the defect is somewhere between the center and the boundary [cf.~Fig.~\ref{Fig:ca}~(a,b)]. Below $c\lowt{a}\hight{s2}$ only moving boundary defect states occur.
The saddle-node bifurcation $c\lowt{a}\hight{s2}$ is also traced in the $(c\lowt{a},c\lowt{hpa})$-plane and is included in Fig.~\ref{Fig:pd-app}. We note that Fig.~\ref{Fig:pd-app} indicates the existence of bifurcations of codimension two. A deeper analysis of the changes occuring there remains a task for the future. Therefore, we only show continuation results directly related to the bistabilities discussed here.


\balance

\providecommand*{\mcitethebibliography}{\thebibliography}
\csname @ifundefined\endcsname{endmcitethebibliography}
{\let\endmcitethebibliography\endthebibliography}{}

\end{document}